\documentclass[12pt]{article}

\usepackage{epstopdf}
\usepackage{amssymb}
\usepackage{times}
\usepackage{graphicx}
\usepackage{color}
\usepackage{subfigure}
\usepackage{jheppub}

\begin{document}

\baselineskip 6mm
\renewcommand{\thefootnote}{\fnsymbol{footnote}}


\newcommand{\nc}{\newcommand}
\newcommand{\rnc}{\renewcommand}


\newcommand{\tcb}{\textcolor{blue}}
\newcommand{\tcr}{\textcolor{red}}
\newcommand{\tcg}{\textcolor{green}}


\def\be{\begin{eqnarray}}
\def\ee{\end{eqnarray}}
\def\nn{\nonumber\\}


\def\ct{\cite}
\def\la{\label}
\def\eq#1{(\ref{#1})}


\def\a{\alpha}
\def\b{\beta}
\def\g{\gamma}
\def\G{\Gamma}
\def\d{\delta}
\def\D{\Delta}
\def\e{\epsilon}
\def\et{\eta}
\def\ph{\phi}
\def\Ph{\Phi}
\def\ps{\psi}
\def\Ps{\Psi}
\def\k{\kappa}
\def\l{\lambda}
\def\L{\Lambda}
\def\m{\mu}
\def\n{\nu}
\def\th{\theta}
\def\Th{\Theta}
\def\r{\rho}
\def\s{\sigma}
\def\S{\Sigma}
\def\ta{\tau}
\def\o{\omega}
\def\O{\Omega}
\def\pr{\prime}


\def\half{\frac{1}{2}}

\def\goto{\rightarrow}

\def\na{\nabla}
\def\grad{\nabla}
\def\curl{\nabla\times}
\def\div{\nabla\cdot}
\def\pa{\partial}
\def\fr{\frac}

\def\bra{\left\langle}
\def\ket{\right\rangle}
\def\lb{\left[}
\def\lc{\left\{}
\def\ls{\left(}
\def\lp{\left.}
\def\rp{\right.}
\def\rb{\right]}
\def\rc{\right\}}
\def\rs{\right)}

\def\vac#1{\mid #1 \rangle}


\def\td#1{\tilde{#1}}
\def\check{ \maltese {\bf Check!}}


\def\Tr{{\rm Tr}\,}
\def\det{{\rm det}}
\def\inf{\infty}


\def\bc#1{\nnindent {\bf $\bullet$ #1} \\ }
\def\ch {$<Check!>$ }
\def\ss {\vspace{1.5cm}}
\def\inf{\infty}

\begin{titlepage}

\hfill\parbox{2cm} { }

 
\vspace{2cm}

\begin{center}
{\Large \bf Holographic RG Flows from Running Couplings}

\vskip 1cm
{Chanyong Park$^{a}$\footnote{e-mail : cyong21@gist.ac.kr} and
Jung Hun Lee$^b$\footnote{e-mail : junghun.lee@jbnu.ac.kr}}

\vskip 1cm

{\it $^a\,$ Department of Physics and Photon Science, Gwangju Institute of Science and Technology,  Gwangju  61005, Korea } \\
{\it  $^b\,$Department of Physics, Jeonbuk National University, Jeonju, 54896, Korea}\\
\end{center}

\thispagestyle{empty}

\vskip1cm


\centerline{\bf ABSTRACT} \vskip 4mm

We investigate the RG flow in the holographic setup and study new IR physics at the IR fixed point, when a relevant scalar operator deforms a two-dimensional conformal field theory (CFT). On the dual gravity side, the renormalization group (RG) flow from the UV to the IR CFT can be described by a scalar field rolling from an unstable to a stable equilibrium point. Specifically, to make the scale dependence of the RG flow more explicit, we study the running coupling constant obtained by solving the beta function. After considering a simple scalar potential, we study the change of the coupling constant and its effect on the ground state along the RG flow. We showed that the central charge monotonically decreases along the RG flow, and its value at the IR fixed point is determined by the conformal dimension of the deformation operator and the coupling constant at the IR fixed point. We further investigate how changing the ground state influences the two-point function and conformal dimension of a local operator in the probe limit.

\vspace{2cm}
 

\end{titlepage}

\renewcommand{\thefootnote}{\arabic{footnote}}
\setcounter{footnote}{0}



\section{Introduction}

Since the holography proposal, there have been many attempts to understand various strongly interacting systems in the dual gravity theory. This holographic relation or the AdS/CFT correspondence is relatively manifest in large-$N$ supersymmetric theories, $N=4$ Super Yang-Mills theory and ABJM theory \cite{Maldacena:1997re, Maldacena:1998im, Witten:1998qj, Witten:1998zw, Gubser:2002tv, Aharony:2008ug, Klebanov:2009sg}. In these cases, infrared (IR) physics is trivial because of conformal symmetry. To investigate nontrivial IR physics holographically, one needs to break the conformal symmetry. In this work, we consider a relevant deformation of a UV conformal field theory (CFT) and explore new macroscopic features near the IR fixed point by applying the holographic method.

Holography is one of the useful techniques for understanding strongly interacting systems \cite{deBoer:1999tgo, Petrini:1999qa, Freedman:1999gp,deHaro:2000vlm,Ryu:2006bv, Ryu:2006ef}. Although the dual gravity theories of maximally supersymmetric gauge theories have been known, finding the dual gravity of non-supersymmetric theories appearing in nuclear and condensed matter physics remains an unresolved issue. Nevertheless, once one identifies the dual gravity theories of real physical systems, one gets more information about the new physics laws governing real systems in the IR region. Since a relevant deformation breaks the conformal symmetry, it usually generates a nontrivial renormalization group (RG) flow and may allow for a new IR CFT. In this case, the appearance of the IR CFT is due to the nonperturbative RG flow, including all quantum corrections. In the traditional quantum field theory (QFT), unfortunately, it is very difficult to analyze such a nonperturbative RG flow. However, one can investigate a nonperturbative RG flow by applying the holographic method  \cite{Bianchi:2001kw, Myers:2010tj, Casini:2015woa, Gukov:2015qea, Kiritsis:2016kog, Ghosh:2017big, Park:2018ebm, Gursoy:2018umf, Ghosh:2018qtg, Park:2019pzo}.

Recently, quantum entanglement entropy has been widely studied in high-energy physics, as well as in quantum information and condensed matter theory \cite{Ryu:2006bv, Ryu:2006ef, Klebanov:2007ws, Karch:2010mn, Takayanagi:2011zk, Ogawa:2011bz, Erdmenger:2015spo, Erdmenger:2015xpq, Rangamani:2016dms, Takayanagi:2017knl}. Entanglement entropy is one of the key quantities characterizing the quantum correlation between subregions and plays the role of an order parameter in quantum phase transitions \cite{Vidal:2002rm, Levin:2006zz}. It is well known that the entanglement entropy in the UV limit shows a universal feature proportional to the area of the entangling surface \cite{Srednicki:1993im, Calabrese:2004eu}. For a two-dimensional CFT, in particular, the entanglement entropy diverges logarithmically, and the coefficient of the logarithmic divergence corresponds to the central charge of the CFT. In holographic studies, it was proposed that the entanglement entropy can be described by the area of a minimal surface extending into the dual geometry \cite{Ryu:2006bv, Ryu:2006ef, Casini:2011kv}. It was also shown that the holographic method correctly reproduces the area law of entanglement entropy in the UV limit. At a phase transition point in the IR region, another logarithmic divergence can occur due to the restoration of conformal symmetry. This is also true at an IR fixed point. To demonstrate this, we consider a three-dimensional Einstein-scalar gravity theory having several local extrema. In this case, a geometric solution interpolating between a local maximum and a minimum describes the RG flow of a dual two-dimensional QFT from a UV to an IR fixed point. On this interpolating geometry, we investigate the RG flow of the entanglement entropy, which is crucially associated with the change of the ground state. After numerically studying the nonperturbative $\beta$-function, $c$-function, and entanglement entropy along the RG flow, we further investigate their analytic structure near the IR fixed point. We show that the entanglement entropy diverges logarithmically at the IR fixed point and that the $c$-function, representing the degrees of freedom, monotonically decreases as the coupling constant increases along the RG flow. In addition, we show that the IR central charge is crucially related to the conformal dimension of the deformation operator and the coupling constant at the IR fixed point.

The main focus of this work is to elucidate how such nonlocal observables along the RG flow can be reorganized and interpreted in terms of a running coupling associated with the turning point of the extremal surface or geodesic probe. In particular, we clarify that this turning point coupling should not be viewed as a quantity determined solely by the IR geometry, nor merely as a reparametrization of known IR CFT behavior. Rather, it depends on the full bulk solution across the entire RG trajectory, thereby encoding nontrivial information about the crossover scale between UV and IR regimes. Within this framework, the familiar logarithmic behavior of entanglement entropy in the IR emerges naturally as a limiting case, while the turning point parametrization provides a unified description of observables even at intermediate scales, where neither the UV nor the IR CFT approximation is fully valid.

We further investigate the correlation function of an additional local operator when the ground state changes along the RG flow. According to the AdS/CFT correspondence, it was proposed that a geodesic length $L(t,x_1;t,x_2)$ connecting two local operators at the boundary is related to a two-point function \cite{Witten:1998qj, Susskind:1998dq, Solodukhin:1998ec, DHoker:1998vkc, Liu:1998ty, Balasubramanian:1999zv, Louko:2000tp, Kraus:2002iv, Fidkowski:2003nf, Park:2020nvo, Rodriguez-Gomez:2021pfh, Park:2024pkt}:\begin{equation}\langle O (t ,x_1) \ O (t , x_2) \rangle \sim e^{- \Delta_{UV} , L(t ,x_1;t ,x_2) /R_{UV}} ,\end{equation}where $\Delta_{UV}$ denotes the conformal dimension of a local operator $O (t ,x_1)$ at the UV fixed point. This holographic prescription describes the two-point function of the dual QFT in the probe limit. The change of the ground state with the variation of a coupling constant can affect such a two-point function by modifying the conformal dimension, which we call the anomalous dimension. We investigate how the change of the ground state leads to the variation of the conformal dimension numerically across the entire energy range and look into the anomalous dimension analytically near the IR fixed point. We also show that the scaling dimension of a local operator monotonically decreases along the RG flow without breaking the unitary bound.

The rest of this paper is organized as follows. In Sec. 2, we consider a dual gravity theory characterized by the RG flow of the dual QFT. We investigate the change of the coupling constant and ground state along the RG flow and determine the IR central charge as a function of only the conformal dimension and coupling constant. In Sec. 3, we study how the change of the ground state affects the two-point function and conformal dimension of a local operator in the probe limit. We conclude this work with some discussions in Sec. 4.


\section{Holographic dual of an RG flow from a UV to IR fixed point}

We take into account a two-dimensional CFT deformed by a relevant operator. Even when a UV theory is weakly interacting, IR physics triggered by a relevant deformation can strongly interact and lead to a new IR fixed point with the scale symmetry restoration \cite{Calabrese:2004eu, Peschel1998DensitymatrixSF, Kim:2016hig, Calabrese:2005zw}. Therefore, knowing IR physics requires understanding a nonperturbative RG flow. To understand such nonperturbative features in the holographic setup, we consider a three-dimensional gravity theory with a bulk scalar field  
\be
S =  \fr{1}{16 \pi G} \int d^3 X \sqrt{-g} \ls {\cal R}  - \half  \pa_M  \ph  \pa^M \ph - \fr{V(\ph)}{R_{UV}^2}   \rs .    \la{Action:sGravity}
\ee  
To describe an RG flow from a UV CFT to a new IR CFT, we introduce the following simple scalar potential as a toy model
\be
V (\ph)  =   2 R_{UV}^2 \L_{UV} + \fr{M_\ph^2}{2} \ph^2 + \fr{\l}{4} \ph^4 =  2 R_{UV}^2 \L_{UV} + \fr{\l}{4} \ph^2 \ls \ph^2 - 2 \fr{m_\ph^2}{\l} \rs ,       \la{Result:scalarpot}
\ee
where $M_\ph^2 = - m_\ph^2 < 0$, $\l >0$ and $\L_{UV}= - 1/R_{UV}^2$ with a UV AdS radius $R_{UV}$. Note that here we exploit dimensionless field $\ph$ and coupling constants, $m_\ph^2$ and $\l$. The scalar potential considered here allows one local maximum, $V = - 2 \L_{UV}$ at $\ph=0$, and two degenerated local minima at $\ph_\pm = \pm m_\ph/\sqrt{\l}$. At the local minima, the value of the scalar potential can be reinterpreted as an IR cosmological constant
\be
V (\ph_\pm) \equiv 2 R_{UV}^2 \L_{IR} =  2 R_{UV}^2 \L_{UV} - \fr{m_\ph^4}{4 \l}  .
\ee
This admits another AdS space in the IR limit, whose AdS radius is reduced to 
\be
R_{IR}  =   \sqrt{  \fr{8 \l  }{8 \l + m_\ph^4}}  \ R_{UV}  .  \la{Result:Rir}
\ee

From now on, we concentrate on the RG flow from a UV CFT to another IR CFT. Such an RG flow in the holographic setup can be realized by a geometric solution that interpolates the local maximum to one of the local minima. Since a local maximum or minimum on the gravity side corresponds to an unstable or stable equilibrium point, respectively, the scalar field generally rolls down from a maximum to a minimum. In other words, a rolling motion in the gravity maps to an RG flow of the dual QFT. From now on, we take into account the RG flow from a UV CFT at $\ph=0$ to an IR CFT at $\ph_{IR} = \ph_+ $. On the dual QFT side, the corresponding RG flow can be triggered by a relevant operator. To consider a relevant operator, we assume that the bulk scalar field has a negative mass squared in the range $-1 < M_\ph^2 \equiv - m_\ph^2  <0$. Suppose that the bulk scalar field depends only on the radial coordinate, so boundary Lorentz symmetry is preserved. In this case, the most general metric ansatz is given by
\be
ds^2 = dy^2 + e^{2 A(y)} \et_{\m\n} dx^\m dx^\n .    \la{Metric:exact}
\ee
Here, UV and IR fixed points appear at $y=\infty$ and $- \infty$, respectively.

A geometric solution interpolating two fixed points is obtained by solving the equations of motion  
\be
0 &=& A'^2 - \fr{1}{4} \ph'^2  - \fr{m_\ph^2}{4 R_{UV}^2} \ph^2 + \fr{\l}{8 R_{UV}^2 }\ph^4 -  \fr{1}{R_{UV}^2} , \\
0 &=&  A'' + A'^2 + \fr{1}{4} \ph'^2 - \fr{m_\ph^2}{4 R_{UV}^2} \ph^2 + \fr{\l}{8 R_{UV}^2 }\ph^4 -  \fr{1}{R_{UV}^2}  , \\
0 &=& \ph'' + 2 A' \ph' + \fr{m_\ph^2}{R_{UV}^2} \ph  - \fr{\l}{ R_{UV}^2 }\ph^3 ,
\ee
where the prime means a derivative with respect to the radial coordinate, $y$. Here, the first equation is a constraint, and the others determine the dynamics of $\ph$ and $A$. After taking $R_{UV}=1$, $m_\ph=\sqrt{3}/2$, $\l = 0.1$ and $G = R_{UV}=1$, we solve the above equations numerically and depict the numerical result in Fig. 1. At the UV fixed point, the result in Fig. 1 shows that the asymptotic geometry is given by an AdS space with $\ph=0$ and $A(y) = y/R_{UV}$ at $y=\infty$. On the other hand, $\ph$ and $A'$ approach $\ph_{IR} = 2.7386$ and $A' =1.3050$ at the IR fixed point ($y = - \infty$) which again leads to another AdS space, $A(y) = y/R_{IR}$ at $y=- \infty$ with a new IR AdS radius $R_{IR} = 0.7663 \, R_{UV}$. In the holographic study, these two UV and IR AdS spaces are mapped to UV and IR fixed points of the dual QFT where a conformal symmetry is restored with vanishing $\b$-functions.

\begin{figure}
\begin{center}
\vspace{-0.5 cm}
\hspace{0.5 cm}
\subfigure[$\ph(y)$]{ \includegraphics[angle=0,width=0.35 \textwidth]{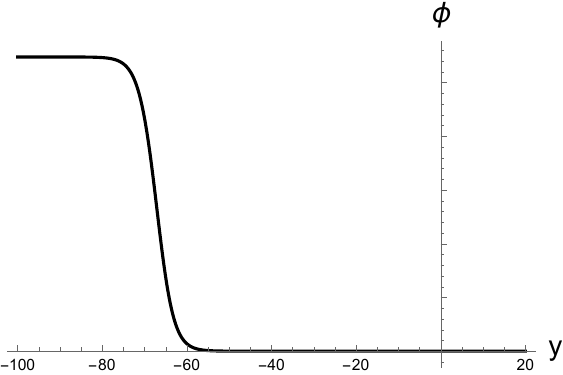}}
\hspace{0.5cm}
\subfigure[$dA(y)/dy$]{ \includegraphics[angle=0,width=0.35 \textwidth]{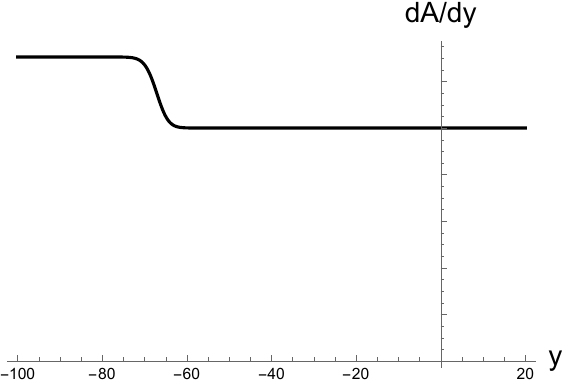}}
\vspace{-0cm}
\caption{Profiles of (a) the scalar field $\ph (y)$ and (b) the derivative of the metric factor, $d A(y) / dy$, where we take $R_{UV}=1$, $m_\ph=\sqrt{3}/2$ and $\l = 0.1$. At the UV fixed point ($y = \infty$), $\ph =0 $ and $d A /dy = 1$, whereas they are modified into $\ph = \ph_{IR}= 2.7386$ and $d A  /dy =1.3050$ at the IR fixed point ($y = -\infty$).}
\end{center}
\end{figure}

When the scalar field is absent, the previous gravity theory allows an AdS space as an exact solution. Since the isometry of an AdS space is equivalent to the conformal symmetry of the dual QFT, a $(d+1)$-dimensional AdS space is mapped to the ground state of a $d$-dimensional CFT. Therefore, the above interpolating geometry can be regarded as the deformation of the CFT. For $0< m_\ph^2 <1$, the bulk field $\ph$ rapidly suppresses in the asymptotic region ($y \to \infty$) and the effect of $ \l \ph^4$ becomes negligible. As a result, the metric factor is well approximated by the AdS one, $A(y) = y/R_{UV}$, which is the dual of a UV CFT. Introducing a new radial coordinate $z$ to describe UV physics
\be
z = R_{UV} \ e^{- y/R_{UV}} ,  \la{Result:UVcoord}
\ee
the asymptotic geometry ($z \to 0$) can be described by 
\be
ds_{UV}^2 = \fr{R_{UV}^2}{z^2} \ls - dt^2 + dx^2 + d z^2\rs .     \la{Result:UVmetric}
\ee
Ignoring $ \l \ph^4$ in the asymptotic region, the scalar field is governed by
\be
0 
=   \pa_z^2 \ph- \fr{1}{z} \  \pa_z  \ph+ \fr{ m_\ph^2 }{z^2}  \ph  ,
\ee
and gives rise to the following profile
\be
\ph (z) = c_1 z^{2-\D_{UV} } + c_2 z^{\D_{UV} } ,    \la{Result:bulkfield}
\ee
where a conformal dimension of a dual scalar operator becomes in the standard quantization
\be
\D_{UV} = 1 + \sqrt{1 - m_\ph^2}  .
\ee
It is worth noting that, since $M_\ph^2$ lies within the unitary range, one can also adopt an alternative quantization, in which the conformal dimension is taken to be $2-\D_{UV}$ rather than $\D_{UV}$ \cite{Klebanov:1999tb}.

Denoting the dual operator of the bulk field $\ph$ by $O_\ph$, it can be classified as relevant, irrelevant, or marginal depending on its conformal dimension \ct{Gukov:2015qea}. This classification is associated with the scalar field's mass and a $\b$-function on the dual QFT side. If the conformal dimension is less than the spacetime dimension, $\D_{UV}< 2$, we call it a relevant operator, which constrains the scalar field mass to satisfy $M_\ph^2 < 0$, which is also related to a negative $\b$-function. To see this,  we identify the value of the scalar field at the boundary with the coupling constant of dual QFT. Then, the previous interpolating geometry results in the following $\b$-function near the UV fixed point
\be
\b_{\ph} = \m \fr{d \ph}{d \m} =  - (2 -\D_{UV} ) \ph + \cdots < 0  ,   \la{Result:betafunph}
\ee
where the ellipsis implies higher-order small quantum corrections. In this case, a negative $\b$-function implies that the coupling constant increases along the RG flow. Solving the $\b$-function equation, we obtain $\ph$ as a function of the RG scale  
\be
\ph \sim \m^{- (2 - \D_{UV})} + \cdots . 
\ee
This determines the scaling behavior of the coupling constant, which is consistent with our previous choice, taking the boundary value of $\ph$ as the coupling constant. Using the scalar field's profile and the interpolating geometry in Fig. 1, the nonperturbative $\b$-function is plotted in Fig. 2(a). Fig. 2(a) reveals that a new IR fixed point appears at $\ph=\ph_{IR}$ with the vanishing $\b$-function, as mentioned before. In general, a two-dimensional CFT can be characterized by the central charge. To further understand the new IR fixed point, we consider a $c$-function describing the RG flow of the central charge. In the holographic setup, a $c$-function is derived by \ct{Deger:2002hv,Gubser:2000nd,Park:2018ebm}
\be
c = \fr{3 }{2 G} \fr{1}{A'} .     \la{Relation: HoloC}
\ee 
In Fig. 2(b), we numerically plot the $c$-function as a function of the coupling constant. The numerical results in Fig. 2 (b) show that the central charge, which reflects the effective degrees of freedom of the system, decreases monotonically as the coupling constant increases along the RG flow. The $c$-function has $c_{UV} = 1.5$ at the UV fixed point, whereas it approaches $c_{IR} = 1.149$ at the IR fixed point. In the following, we examine the origin of this specific value at the IR fixed point.

\begin{figure}
\begin{center}
\vspace{-0.5 cm}
\hspace{0.5 cm}
\subfigure[$\b$-function]{\includegraphics[angle=0,width=0.4 \textwidth]{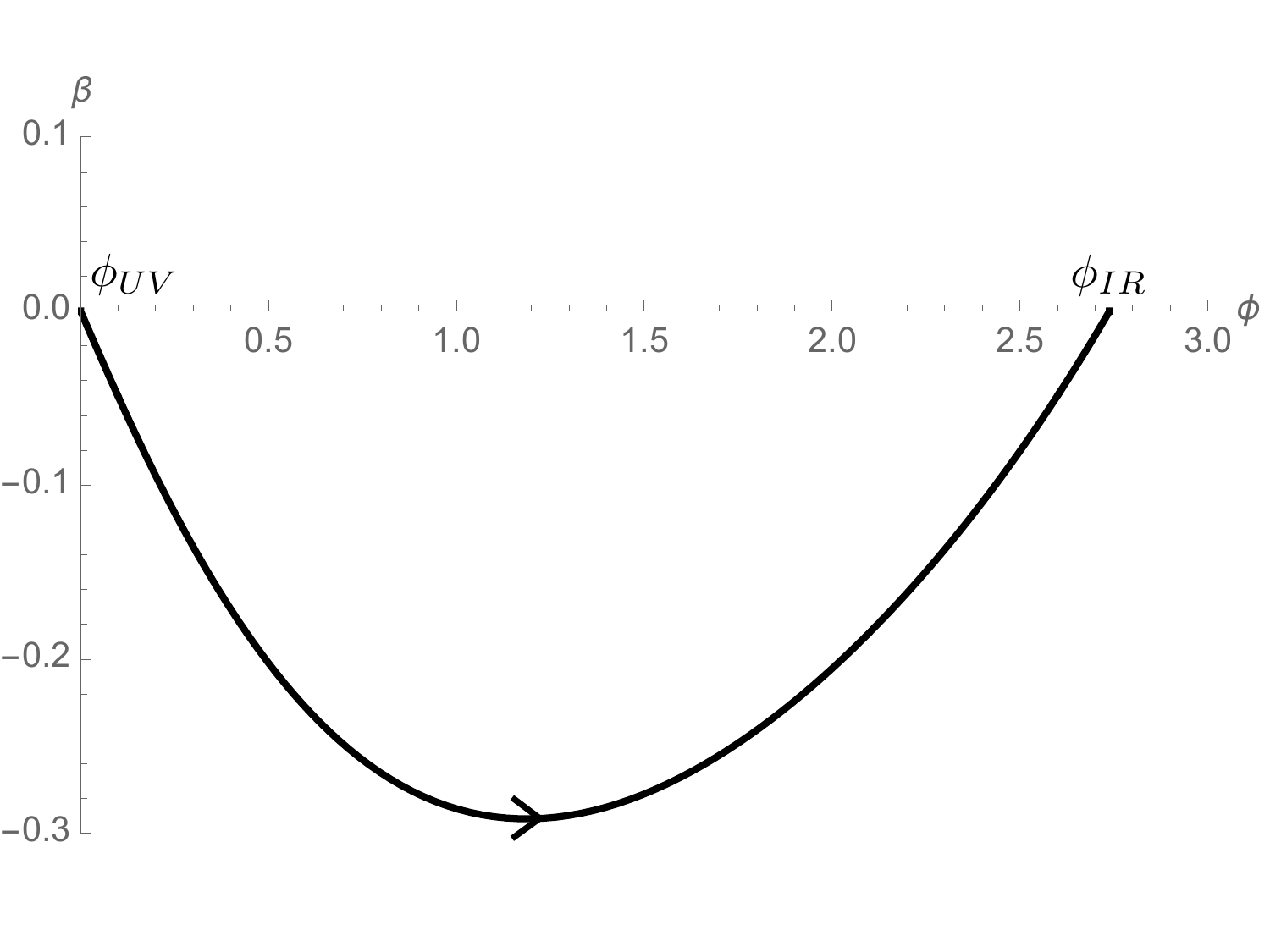}}
\hspace{0.5cm}
\subfigure[$c$-function]{ \includegraphics[angle=0,width=0.45\textwidth]{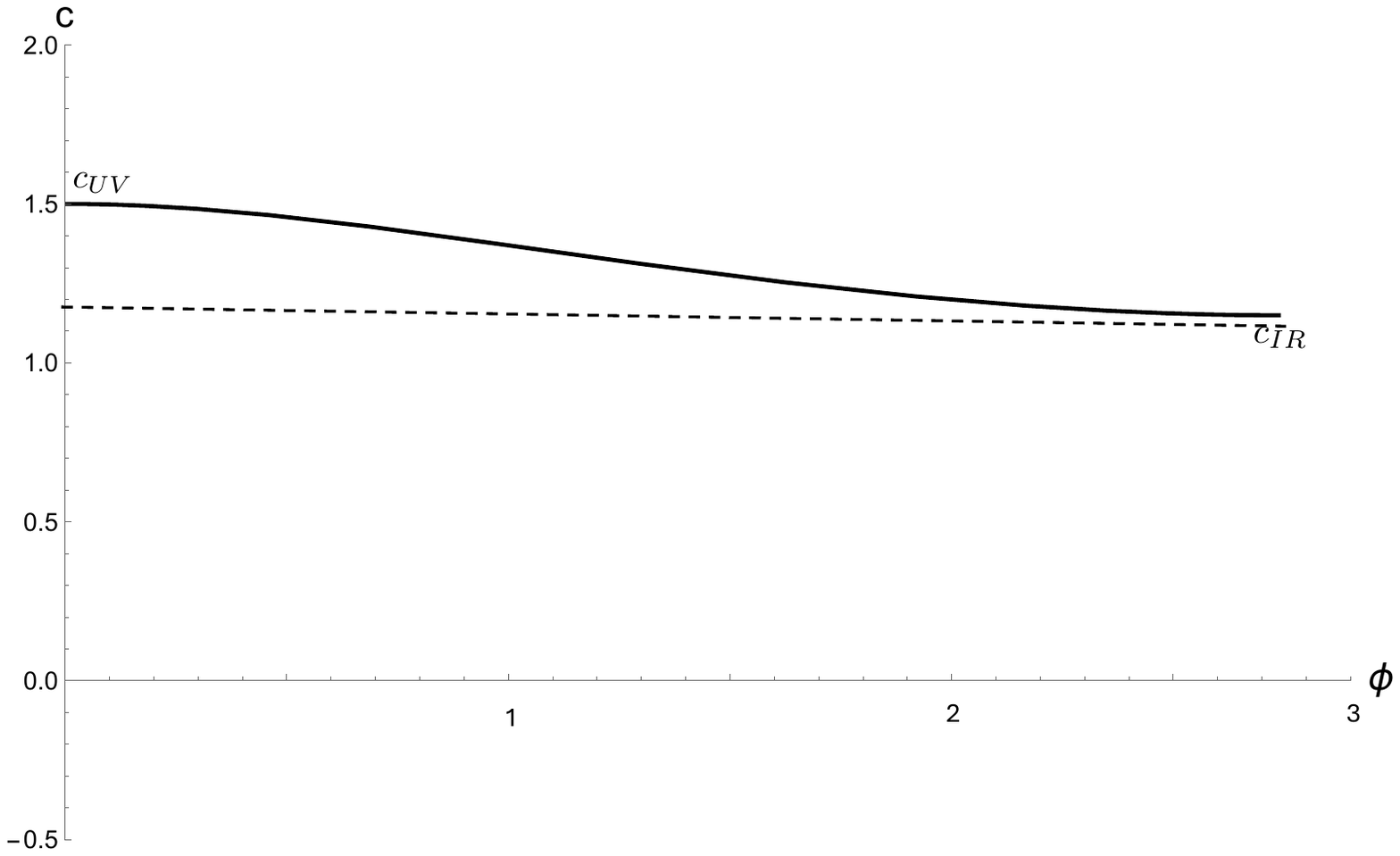}}
\vspace{0.cm}
\caption{ From the previous interpolating geometry, we derive nonperturbative (a) $\b$-function  and  (b) $c$-function of the dual QFT, where we take $m_\ph=\sqrt{3}/2$, $\l = 0.1$ and  $G=R_{UV} =1$.  }
\end{center}
\end{figure}

To understand the variation of a ground state along the RG flow and properties of the IR physics, we look into the RG flow of the entanglement entropy. In the holographic setup, the entanglement entropy is determined by the area of a minimal surface extending to the dual geometry. In the previous interpolating geometry \eq{Metric:exact}, the entanglement entropy is governed by  \cite{Ryu:2006bv,Ryu:2006ef}
\be
S_E = \fr{1}{4 G} \int_{- \ell/2}^{\ell/2} dx \sqrt{ y'^2 + e^{2 A(y) }} ,
\ee
where $\ell$ and the prime denote the size of a subsystem and a derivative with respect to $x$. Using the conserved quantity, the subsystem size and entanglement entropy can be reexpressed as functions of a turning point $y_t$, at which $y'=0$,
\be
\ell &=& \int_{y_t}^{\L_y} dy \ \fr{2}{e^{2 A (y)} \sqrt{e^{- 2 A_t} - e^{- 2 A  (y)}}} , \\
S_E &=& \fr{1}{2 G}\int_{y_t}^{\L_y} dy \ \fr{ e^{- A_t} }{ 
\sqrt{e^{- 2 A_t} - e^{- 2 A (y)}}}  ,  \la{Relation: HEEy}
\ee
where $A_t = A(y_t)$ and $\L_y$ denotes an appropriate UV cutoff. Recalling that the subsystem size is closely related to the turning point, the entanglement entropy is determined as a function of the turning point, which plays the role of the RG scale. In addition, the value of the scalar field at the turning point, $\ph_t = \ph(y_t)$, is identified with the coupling constant measured at the RG scale. Combining them, we finally obtain the entanglement entropy relying on the coupling constant in Fig. 3(a). The numerical result in Fig. 3(a) shows that the entanglement entropy increases logarithmically at the UV and IR fixed points ($\ph_t=0$ and $ \ph_t= 2.7386$). This is because of the restoration of conformal symmetry at the fixed points.

\begin{figure}
	\begin{center}
		\vspace{-0.5 cm}
		\hspace{0.5 cm}
		\subfigure[]{ \includegraphics[angle=0,width=0.34 \textwidth]{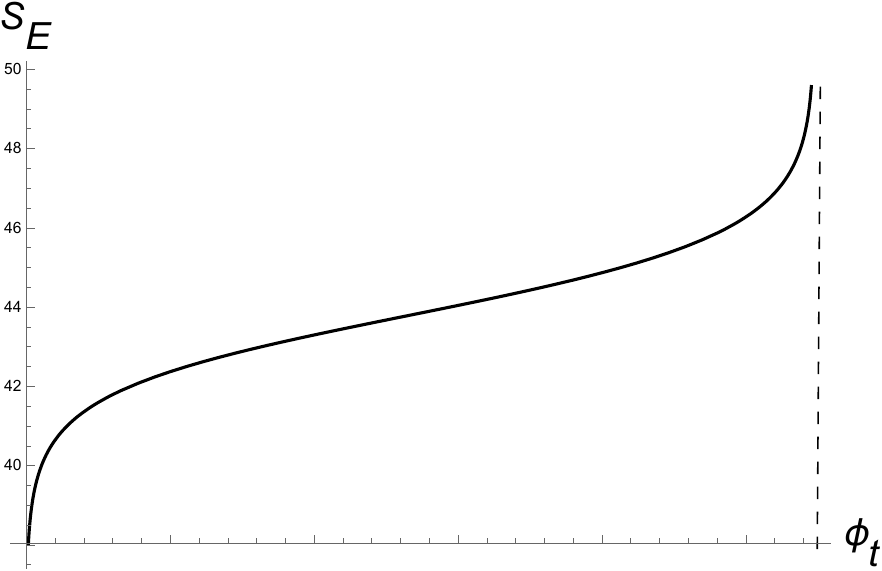}}
		\hspace{0.5cm}
		\subfigure[]{ \includegraphics[angle=0,width=0.45\textwidth]{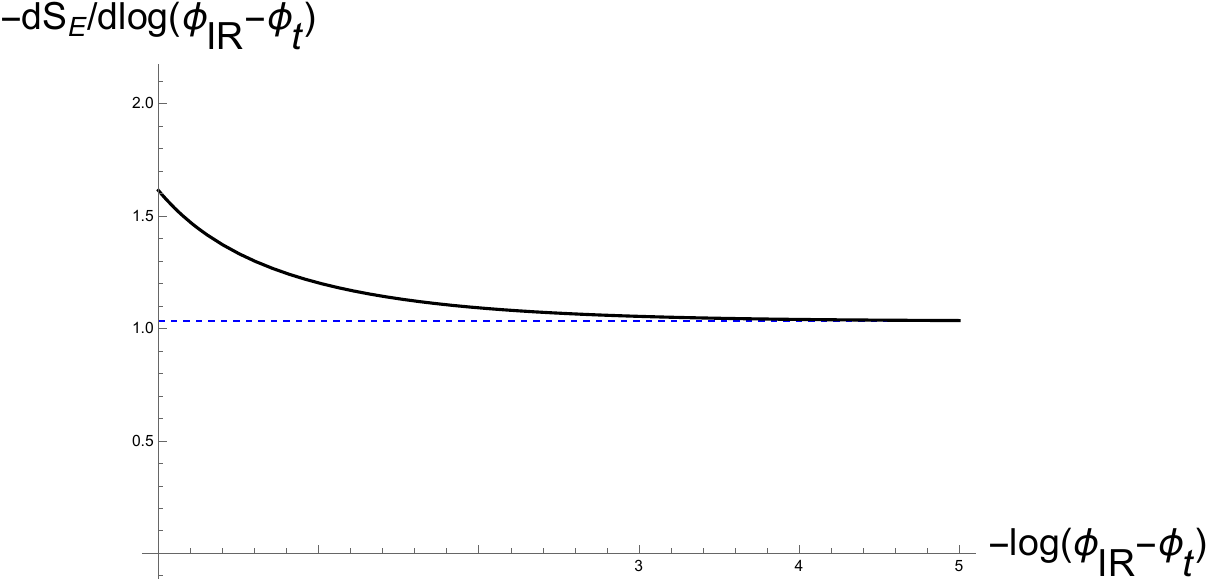}}
		\vspace{-0cm}
\caption{(a) Entanglement entropy depending on the coupling constant and (b) the coefficient of the IR entanglement entropy where we fix $G=1$. (a) The entanglement entropy shows a logarithmic divergence at $\ph_t \to \ph_{IR}$. (b) When $-\log(\ph_{IR} -\ph_t)  \to \infty$ near the IR fixed point, $- d S_E/ d \log(\ph_{IR} -\ph_t)$ approaches $1.0316$ which is consistent with the value expected by \eq{Result:IREE}.}
\end{center}
\end{figure}

Although the entanglement entropy was calculated numerically in Fig. 3, we need to derive an analytic expression to understand the coupling dependence of the entanglement entropy. Recalling that the turning point determines the RG scale as $\m \sim 1/ z_t \sim 1/ \ell$,  the turning point $z_t$ in the UV regime is related to the coupling constant $\ph(z_t)= \ph_t$ 
\be
z_t \approx \ls \fr{\ph_t}{c_1} \rs^{1/(2 -\D_{UV})} .
\ee
Therefore, the UV entanglement entropy becomes  
\be
S_E  =  \fr{ R_{UV}}{2  \ (2 -\D_{UV})  G } \log \fr{\ph_t}{\ph_{\e} } + {\cal O} \ls \ph_t^{1/(2 -\D_{UV})}  \fr{}{} \rs ,
\ee
where $\ph_{\e} =\ph(\e)$ means the coupling constant at the UV fixed point. In terms of the entanglement entropy, a $c$-function is defined as
\be
c = - 3 \fr{d S_E}{d \log \m}   .
\ee 
Therefore, the central charge near the UV fixed point is reduced to
\be
c_{UV} = - 3 \fr{d S_E}{d \log \m} =  3 \fr{d S_E}{d \log \ell} = \fr{3 R_{UV}}{2 G} .
\ee 
This provides the consistent result with that derived from \eq{Relation: HoloC}.

Now, let us focus on the entanglement entropy for a large subsystem size ($y_t \to - \infty$). The IR entanglement entropy is governed by the integration over the entire $z$ in \eq{Relation: HEEy}. Nonetheless, the leading contribution is dominated by the minimal surface extending to the IR region. Introducing a new coordinate to describe the IR geometry for $y \to - \infty$
\be
\bar{z} = R_{IR} \ e^{- y/R_{IR}}  ,      \la{Result:IRcoord}
\ee
the IR AdS metric is written as
\be
ds_{IR}^2 = \fr{R_{IR}^2}{\bar{z}^2} \ls - d \bar{t}^2 + d \bar{x}^2 + d\bar{z}^2\rs , \la{Result:IRmetric}
\ee
Near the IR fixed point, the scalar field $\ph = \ph_{IR} + \d \ph$ can be characterized by a small fluctuation $\d \ph$ around $\ph_{IR}$, which satisfies the following equation of motion
\be
0   =   \pa_{\bar{z}}^2  \d \ph- \fr{1}{\bar{z}} \  \pa_{\bar{z}}   \d \ph -  \fr{2 \ m_\ph^2}{\bar{z}^2} \fr{R_{IR}^2}{R_{UV}^2} \ \d \ph  .
\ee
The solution of this equation reads
\be
\d \ph = d_1 \,  \bar{z}^{ 2+\d_{IR}  }  -  d_2 \, \bar{z}^{ - \d_{IR}}   ,  
\la{Solution:deltaph}
\ee
where $d_1$ and $d_2$ are two integration constants and $\d_{IR}$ is given by
\be
\d_{IR} 
=  \fr{\sqrt{8 \l + m_\ph^4 +  16 m_\ph^2 \l} }{\sqrt{8 \l + m_\ph^4 }} - 1 .
\ee
As $\bar{z} \to \infty$, the first term of $\d \ph$ diverges for $\d_{IR} > 0$. However, since we assumed at the early stage that $\ph = \ph_{IR}$ at the IR fixed point is finite, the first term in \eq{Solution:deltaph} must vanish. This requires $d_1 =0$. Therefore, the resulting profile of $\ph$ near the IR fixed point is reduced to
\be
\ph (\bar{z}) = \ph_{IR} - d_2 \, \bar{z}^{- \d_{IR}} + \cdots ,  \la{Result:phprofielIR}
\ee 
where the remaining integration constant $d_2$ is associated with $c_1$ and $c_2$ in \eq{Result:bulkfield}. Exploiting this scalar field profile, the IR RG scale is connected to the coupling constant
\be
\m \sim \fr{1}{\bar{z}} \approx  \ls \fr{\ph_{IR} - \ph (\bar{z}) }{d_2} \rs^{1/\d_{IR}}   .
\ee

For $\ell \to \infty$, the entanglement entropy defined on the above IR AdS geometry provides a leading contribution  
\be
S_E \approx \lim_{\ell \to \infty} \fr{R_{IR}}{4 G} \int_{-\ell/2}^{\ell/2} dx   \ \fr{ \sqrt{ \bar{z}'^2 + 1 }}{\bar{z}} .
\ee
After some calculation, the dominant IR entanglement entropy in terms of the coupling constant is given by
\be
S_E  &\approx& 
 - \fr{ \sqrt{8 \, \l} \,  R_{UV} }{ 2  \,    \ls   \sqrt{8 \l + m_\ph^4 +  16 m_\ph^2 \l}   -\sqrt{8 \l + m_\ph^4 } \rs  G }   \log \ls \fr{\ph_{IR} - \bar{\ph}_t }{d_2} \rs   .  \la{Result:IREE}
\ee
where $\bar{\ph}_t = \ph (\bar{z}_t)$ is the coupling constant in the IR regime. This result shows that a new logarithmic divergence appears as the coupling constant approaches the IR value, as mentioned before. Moreover, the IR central charge reads  
\be
 c_{IR} &=&  \fr{3  R_{IR}}{2 G}  
=   \fr{ \sqrt{8}  }{ \sqrt{8 + \D_{UV} (2 - \D_{UV})  \,   \ph_{IR}^2 } }  \  c_{UV},
\ee 
which coincides with the numerical result in Fig. 2(b). Intriguingly, this result reveals that the IR central charge can be uniquely determined by the coupling constant at the IR fixed point, $\phi_{IR}$, and the conformal dimension of the deformation operator at the UV fixed point, $\Delta_{UV}$.

\section{Anomalous dimension of a local operator in the probe limit}

In the previous section, we studied how to describe the nonperturbative RG flow of the ground state. The evolution of the coupling constant along the RG flow modifies the ground state, consequently altering both the entanglement entropy and the central charge. We now investigate how a change of the ground state can affect correlation functions of local operators. To do so, we introduce new local operators, $O_\chi (t_1, x_1)$ and $O_\chi (t_2, x_2)$, into the ground state deformed by $O_\ph$. If the interaction between $O_\chi$ and $O_\ph$ is strong, the effect of $O_\chi$ further modifies the ground state and can spoil the IR fixed point. From now on, we assume that the interaction between local operators is weak during the RG flow and consider a probe limit. Assuming that the scalar operator $O_\chi$ has a conformal dimension $\D_{UV}^\chi$ at the UV fixed point, conformal symmetry constrains its two-point function to be
\be
\bra O_\chi (t_1,x_1) \ O_\chi (t_2, x_2) \ket = \fr{1}{ (-  | t_1 - t_2 |^2 + |x_1 - x_2|^2 )^{\D_{UV}^\chi}}  .
\ee
At the IR fixed point, the restoration of the conformal symmetry again leads to a similar two-point function with a different IR conformal dimension because nonperturbative quantum effects modify the conformal dimension of $O_\chi$ during the RG flow. In this section, we investigate with the aid of holography how the change of the ground state modifies the conformal dimension of $O_\chi$ in the probe limit.

In the holographic setup, it was proposed that a two-point function is described by a geodesic length  connecting two local operators at the boundary \cite{Susskind:1998dq, Solodukhin:1998ec, DHoker:1998vkc, Liu:1998ty, Balasubramanian:1999zv, Park:2020nvo,Park:2020xho, Park:2022mxj, Park:2022abi, Kim:2023fbr}
\be
\bra O_\chi (t ,x_1) \ O_\chi (t , x_2) \ket  \sim e^{- \D_{UV}^\chi  \,  L(t ,x_1;t ,x_2) /R_{UV}} ,   \la{Formula:EET}
\ee
where $ L(t ,x_1;t ,x_2)$ indicates a geodesic length. Before discussing holographic two-point functions, 

On the previous interpolating geometry in Fig. 1(a), the geodesic length is governed by
\be
L(t,x_1;t,x_2) &=& \int_{x_1}^{x_2} dx \  \sqrt{ y'^2 + e^{2 A (y)}}   .
\ee
This, together with the conservation, enables us to represent an operator's distance and geodesic length in terms of the turning point
\be
| x_1 - x_2 | &=&  \int_{y_t}^{\L_y} dy \ \fr{ 2 \,  e^{- 2 A (y)} }{  \sqrt{e^{- 2 A_t} - e^{- 2 A (y)}}}  ,  \nn 
L(t,x_1;t,x_2)   &=& \int_{y_t}^{\L_y} dy \ \fr{ 2 \,  e^{- A_t} }{  \sqrt{e^{- 2 A_t} - e^{- 2 A (y)}}} ,
\ee
where $A_t = A(y_t)$. To study the change in the conformal dimension, we define an effective conformal dimension as
\be
\D_{eff} (|x_1 - x_2|) \equiv - \fr{1}{2} \fr{d \log \bra O_\chi (t,x_1) \ O_\chi (t, x_2) \ket }{d \log |x_1 - x_2| }  .
\ee 
The effective conformal dimension defined here is generally given by a function of the operator's distance, whereas it becomes a constant at fixed points due to scale symmetry. We plot the effective potential numerically in Fig. 4, where we take $\D^{\chi}_{UV} = 1$ for covenience. The numerical simulation shows that the effective conformal dimension monotonically decreases and approaches $\D^{\chi}_{IR} = 0.7663$ at the IR fixed point. This implies that the IR conformal dimension is given by 
\be
\D^{\chi}_{IR}  = 0.7663 \, \D^{\chi}_{UV}  .   \la{Result:IRcondimchi11}
\ee

\begin{figure}
	\begin{center}
		\vspace{-0.5 cm}
		\hspace{0.5 cm}
		\subfigure{\includegraphics[angle=0,width=0.45 \textwidth]{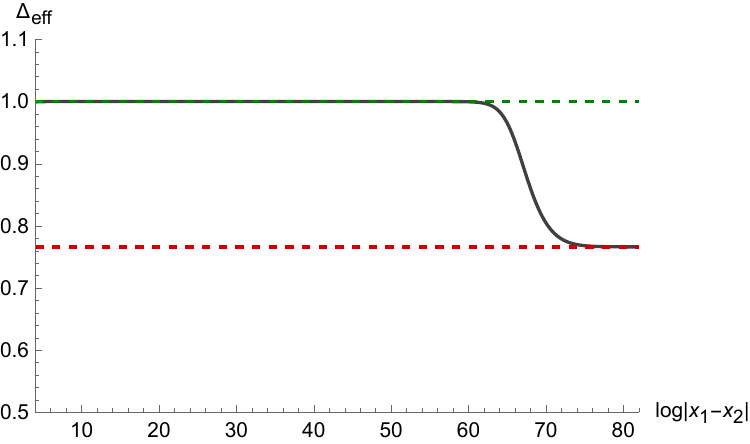}}
		\vspace{-0.0cm}
		\caption{ The RG flow of a conformal dimension of $O_\chi$. When the ground state is deformed by $O_\ph$, the change of the ground state affects another operator $O_\chi$. Assuming that the UV conformal dimension is $\D^\chi_{UV}=1$, the conformal dimension along the RG flow changes into $\D^\chi_{IR}=0.7663$ at the IR fixed point. This is consistent with the previous theoretical expectation in \eq{Result:IRcondimchi}.    }
	\end{center}
\end{figure}

To understand \eq{Result:IRcondimchi11} analytically, we introduce a bulk field $\chi$ as the dual of $O_\chi$. To realize the UV conformal dimension $\D_{UV}^\chi$, we assume that the mass of $\chi$ is given by $- m_\chi^2/R_{UV}^2$ in the asymptotic region. Then, the bulk field $\chi$ in the probe limit satisfies the following linear differential equation  in the asymptotic region
\be
0 
&=&   \pa_z^2 \chi -  \fr{1}{z} \ \pa_z  \chi +  \fr{ m_\chi^2}{z^2}  \chi  .    \la{Equation:UVscalar}
\ee
This allows the following perturbative solution in the asymptotic region
\be
\chi = c_1 \,  z^{2-\D_{UV}^{\chi}} +   c_2 \,  z^{\D_{UV}^{\chi}} ,    \la{Solution:UVchi}
\ee
where $c_1$ and $c_2$ are two integration constants and the dual operator $O_\chi$ has a conformal dimension
\be
\D_{UV}^\chi = 1 + \sqrt{1 - m_\chi^2  }  .
\ee
Here, we should take $m_\chi^2 =1$ to obtain $\D^{\chi}_{UV} = 1$. Using this conformal dimension, the two-point function of $O_\chi$ at the UV fixed point results in
\be
\bra O_\chi (t,x_1) \ O_\chi (t, x_2) \ket = \fr{1}{  |x_1 - x_2|^{2 \D_{UV}^{\chi}}}  .
\ee

In the IR regime, $\chi$ is governed by the following equation
\be
0  &=&     \pa_{\bar{z}}^2 \chi + \ls 1 -   \fr{2 R_{IR}}{R_{UV}} \rs\fr{1}{ \bar{z}} \  \pa_{\bar{z}}  \chi+ \fr{R_{IR}^2}{R_{UV}^2}  \fr{ m_\chi^2 }{\bar{z}^2}  \chi  
\ee 
Solving this equation, the profile of $\chi$ reads  
\be
\chi  = \bar{c}_1 \   \bar{z}^{2  R_{IR} /R_{UV}   - \D_{IR}^\chi  }  +  \bar{c}_2 \  \bar{z}^{   \D_{IR}^\chi },
\ee
where the IR conformal dimension is given by
\be
\D_{IR}^\chi = \fr{R_{IR}}{R_{UV}} \  \D_{UV}^{\chi} =  \fr{\sqrt{  8 \l  } }{ \sqrt{ 8 \l + m_\ph^4 } }  \ \D_{UV}^\chi .    \la{Result:IRcondimchi}
\ee
This indicates that the IR correlator of $\chi$ is reduced to
\be
\bra O_\chi (t,x_1) \ O_\chi (t, x_2) \ket = \fr{1}{  |x_1 - x_2|^{2 \D_{IR}^\chi }}  .
\ee
As a result, the anomalous dimension of $O_\chi$ triggered by the change of the ground state results in
\be
\g_\chi \equiv  \D_{IR}^\chi - \D_{UV}^{\chi}= -  \ls 1 -  \fr{ \sqrt{8}  }{ \sqrt{8 + \D_{UV} (2 - \D_{UV})  \,   \ph_{IR}^2 } }    \rs \,   \D_{UV}^{\chi}     < 0 .
\ee
This analytic result is consistent with the numerical value in \eq{Result:IRcondimchi11}. This result shows that the conformal dimension of a local operator in the probe limit decreases along the RG flow. Nevertheless, such an RG flow does not break the unitary bound, $\D_{IR}^\chi >0$.


\section{Discussion}

Using the holographic method, we studied the RG flow of a two-dimensional QFT from a UV to an IR fixed point. To describe these flows, we considered a UV CFT deformed by a relevant operator. In the dual gravity theory, such an RG flow is represented by a geometric solution smoothly interpolating between a local maximum and a minimum of the scalar potential. Specifically, an unstable local maximum corresponds to the UV fixed point. Near this maximum, the geometry is well-approximated by an AdS space, reflecting the restoration of conformal symmetry. Conversely, the IR fixed point maps to a stable local minimum, where another AdS solution emerges. To demonstrate these features, we introduced a simple scalar potential as a toy model. We then numerically determined the interpolating geometry and explicitly showed that the coupling constant derived from this geometry results in a vanishing $\beta$-function at both the UV and IR fixed points. These numerical results confirm that the RG flow can be visualized as a scalar field rolling from an unstable to a stable equilibrium point in the bulk gravity theory.

In general, a nonperturbative RG flow modifies both the coupling constants and the ground state of a QFT, also altering the structure of quantum correlations. Consequently, the entanglement entropy of the ground state is expected to evolve nontrivially along the RG flow. Using an interpolating geometry, we first investigated the relationship between the RG scale and the coupling constant. We observed that relevant deformations drive the coupling constant to increase toward the IR, which leads to the $c$-theorem. To check this, we analyzed the RG flow of entanglement entropy. By identifying the bulk scalar field with the dual coupling constant, we expressed the entanglement entropy as a function of the coupling. Our numerical results demonstrate that the entanglement entropy diverges logarithmically at both the UV and IR fixed points, reflecting the restoration of conformal symmetry. To gain deeper insight, we also examined the analytic behavior of the entanglement entropy near the IR fixed points. Through perturbative expansions around $\phi=\phi_{IR}$ (IR), we showed that the IR $c$-function is determined by the conformal dimension of the deformation operator and the coupling constant at the IR fixed point, and we also checked that our analytic formula is consistent with the previous numerical result.

Furthermore, we investigated the holographic RG flow of the conformal dimension. Within the holographic framework, a two-point function in the probe limit can be described by the geodesic length connecting two local operators at the boundary. We showed that even in the probe limit, the RG flow of the ground state significantly influences the correlation functions. Our numerical results reveal that the conformal dimension of a local operator decreases monotonically without breaking the unitary bound, as the background coupling constant increases along the RG flow. To further elucidate this behavior, we examined the analytic properties of the two-point function near the IR fixed points and evaluated the anomalous dimension. We found that the analytical results are in perfect agreement with our numerical results, providing a consistent picture for the RG flow of the conformal dimension. 

Although it is not straightforward to establish an exact relation between the entangling subregion and the RG scale, we expect that the reinterpretation method using a coupling constant may help resolve this issue and provide a unified framework for understanding the entire RG sector, including intermediate energy scales.

It would be interesting to investigate the nonperturbative RG flow of the correlation functions beyond the probe limit. We hope to report more interesting results in future work.

\vspace{0.5cm}

{\bf Acknowledgement}

C. P. was supported by the National Research Foundation of Korea (NRF) grant funded by the Korea government (MSIT) (No. RS-2026-25473667).   J. H. L. was supported by the National Research Foundation of Korea grant funded by the Korean government (No. NRF-2021R1C1C2008737)




%

\end{document}